\documentclass[twocolumn,showpacs,preprintnumbers,amsmath,amssymb,prl,aps]{revtex4-2}
\usepackage{graphicx}
\usepackage{natbib}

\usepackage{dcolumn}
\usepackage{amsmath}
\usepackage{braket}
\usepackage{float}
\usepackage{bm}
\usepackage[english]{babel}

\makeatletter
\def\btt#1{\texttt{\@backslashchar#1}}
\DeclareRobustCommand\bblash{\btt{\@backslashchar}}
\makeatother

\usepackage{color}

\begin{document}

\preprint{HEP/123-qed}

%%%%%%%%%%%%%%%%%%%%%%%%%%%%%%%%%%%%%%%%%%%%%%%%%%%%%%%%%%%%%%%%%%%%%%%%%%%%%%%

\title[Short Title]{Microwave hinge states in a simple-cubic-lattice photonic crystal insulator}

\author{Shun Takahashi$^1$}
\author{Yuya Ashida$^1$}
\author{Huyen Thanh Phan$^2$}
\author{Kenichi Yamashita$^1$}
\author{Tetsuya Ueda$^1$}
\author{Katsunori Wakabayashi$^2$}
\author{Satoshi Iwamoto$^{3, 4}$}

\affiliation{
$^1$Kyoto Institute of Technology, Matsugasaki, Sakyo-ku, Kyoto 606-8585, Japan\\
$^2$School of Engineering, Kwansei Gakuin University, Sanda, Hyogo 669-1337, Japan\\
$^3$Research Center for Advanced Science and Technology, The University of Tokyo, 4-6-1 Komaba, Meguro-ku, Tokyo 153-8505, Japan\\
$^4$Institute of Industrial Science, The University of Tokyo, 4-6-1 Komaba, Meguro-ku, Tokyo 153-8505, Japan\\
}

%%%%%%%%%%%%%%%%%%%%%%%%%%%%%%%%%%%%%%%%%%%%%%%%%%%%%%%%%%%%%%%%%%%%%%%%%%%%%%%

\date{\today}

%%%%%%%%%%%%%%%%%%%%%%%%%%%%%%%%%%%%%%%%%%%%%%%%%%%%%%%%%%%%%%%%%%%%%%%%%%%%%%%

\begin{abstract}
We numerically and experimentally demonstrated a higher-order topological state in a three-dimensional (3D) photonic crystal (PhC) with a complete photonic bandgap. 
Two types of cubic lattices were designed with different topological invariants, which were theoretically and numerically confirmed by the finite difference of their Zak phases. 
Topological boundary states in the two-dimensional interfaces and hinge states in the one-dimensional corners were formed according to the higher-order of bulk-boundary correspondence. 
Microwave measurements of the fabricated 3D PhC containing two boundaries and one corner showed a localized intensity, which confirmed the boundary and hinge states.
\end{abstract}

%%%%%%%%%%%%%%%%%%%%%%%%%%%%%%%%%%%%%%%%%%%%%%%%%%%%%%%%%%%%%%%%%%%%%%%%%%%%%%%

\maketitle

%%%%%%%%%%%%%%%%%%%%%%%%%%%%%%%%%%%%%%%%%%%%%%%%%%%%%%%%%%%%%%%%%%%%%%%%%%%%%%%

\section{Introduction}

%%%%%%%%%%%%%%%%%%%%%%%%%%%%%%%%%%%%%%%%%%%%%%%%%%%%%%%%%%%%%%%%%%%%%%%%%%%%%%%

The concept of topology has been applied to photonics \cite{Haldane}, creating the field of topological photonics \cite{Lu,Ozawa,Ota_review,Iwamoto}, which has been intensively studied. 
One-way topological waveguides for microwave and infrared light \cite{Wang,Bahari} and topological lasers \cite{Bahari,Bandres} have been demonstrated. 
In addition to conventional topological edge states based on the bulk-edge correspondence \cite{Hatsugai} where $d$-dimensional topological insulators induce ($d$-1)-dimensional boundary states, higher-order topological insulators providing ($d$-2)- or ($d$-3)-dimensional boundary states have been discussed. 
In particular, in photonic systems, topologically protected zero-dimensional (0D) cavity states in two-dimensional (2D) systems have been experimentally demonstrated for microwaves \cite{Chen} and infrared light \cite{Ota}. 
Such higher-order topological states can be formed in crystalline structures \cite{Xie} whose unit cell shows nonzero bulk polarization \cite{Benalcazar} or a Wannier center with an offset from the origin \cite{Ezawa}, even when the Chern number is zero \cite{Liu}. 

%%%%%%%%%%%%%%%%%%%%%%%%%%%%%%%%%%%%%%%%%%%%%%%%%%%%%%%%%%%%%%%%%%%%%%%%%%%%%%%

Three-dimensional (3D) systems can provide various topological phenomena, as the third dimension expands the opportunities for the structural design of photonic crystals (PhCs) or metamaterials. 
For example, 2D topological surface states have been demonstrated in the microwave regime using Fermi arcs derived from Weyl points \cite{CTChan,Yang}. 
In terms of higher-order topological states, second-order topological states (hinge states) as one-dimensional (1D) waveguides and third-order topological states as 0D cavities have been studied, mainly for acoustic waves \cite{Wei1,Qiu,Xia,Xue2,Weiner,Zhang2}. 
There have been a few reports on microwave hinge states, but only for semimetals without photonic bandgaps \cite{microwave_hinge_PRB,microwave_hinge_OL,microwave_hinge_arXiv}. 
These 3D structures for both acoustic waves and microwaves are complicated and difficult to miniaturize for practical applications in infrared light. 

%%%%%%%%%%%%%%%%%%%%%%%%%%%%%%%%%%%%%%%%%%%%%%%%%%%%%%%%%%%%%%%%%%%%%%%%%%%%%%%

In this study, based on our theoretical and numerical investigations, we fabricate an all-dielectric 3D PhC insulator composed of simple cubic lattices on a millimeter scale and measure the hinge state in the microwave regime. 
The hinge state is based on a higher-order of bulk-boundary correspondence \cite{Benalcazar,Weiner,Zhang2} with a topological invariant of Zak phases \cite{Zak} in two orthogonal directions \cite{Xie,Liu,Liu2}. 
Owing to the complete photonic bandgap (cPBG) of the 3D PhC, the measured hinge state is isolated from the bulk states. 
Such a hinge state, which is strongly confined by the cPBG can be applied to a low-loss transmission line for microwaves. 
In addition, by scaling the studied PhCs down to a sub-$\mu$m period, hinge states for infrared light using 3D nano-fabrication techniques, such as a micro-manipulation method \cite{Aoki,Aniwat,TakahashiEL,TakahashiAPEX}, can be realized. 

%%%%%%%%%%%%%%%%%%%%%%%%%%%%%%%%%%%%%%%%%%%%%%%%%%%%%%%%%%%%%%%%%%%%%%%%%%%%%%%

\section{Sample design and numerical calculations}

%%%%%%%%%%%%%%%%%%%%%%%%%%%%%%%%%%%%%%%%%%%%%%%%%%%%%%%%%%%%%%%%%%%%%%%%%%%%%%%

\begin{figure}
\centering
\includegraphics[width=7cm]{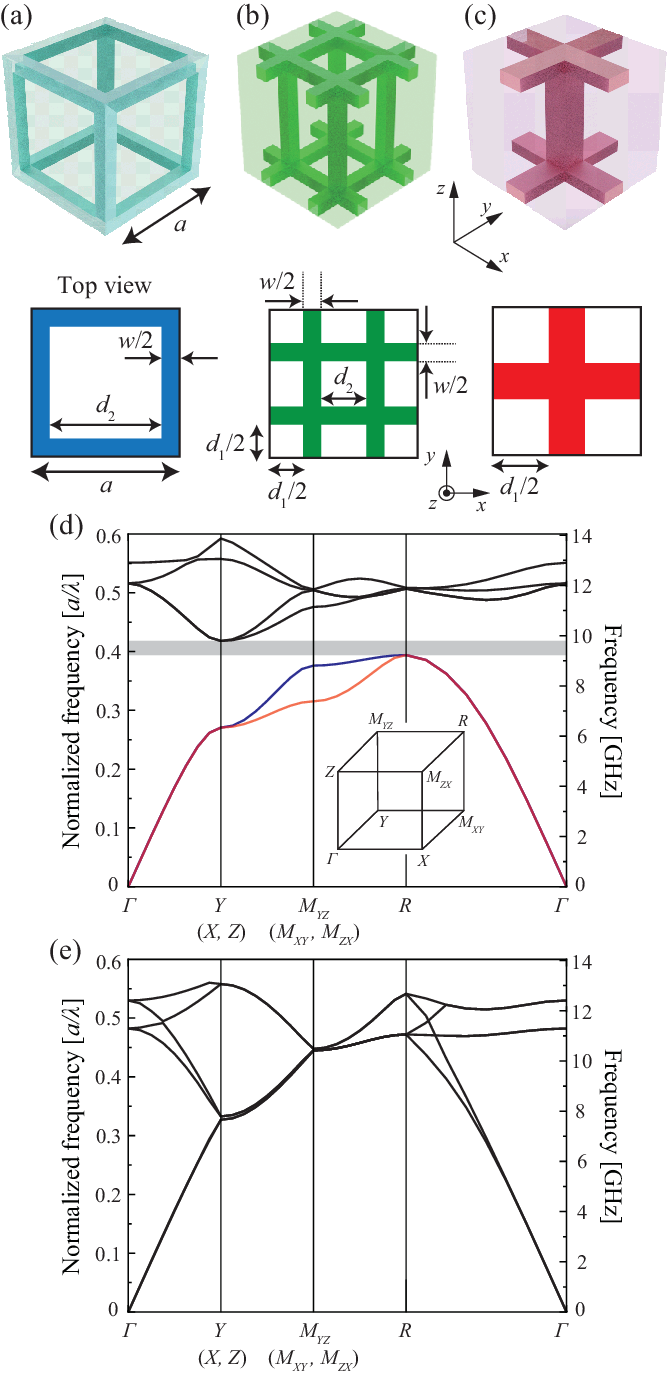}
\caption{
(a)-(c)
Schematics of three kinds of simple-cubic-lattice unit cells. 
These structures are related each other through the variable lengths of $d_1$ and $d_2$, such that ($d_1$, $d_2$) = (0, $a-w$) in (a) and ($d_1$, $d_2$) = ($a-w$, 0) in (c). 
(d)
Numerically obtained photonic band structure of the unit cells in (a) and (c) for the lowest six bands. 
The cPBG is indicated by the shaded region at around a normalized frequency of 0.41, corresponding to 9.5 GHz for the structure with $a$ = 12.8 mm. 
The two colored bands below the cPBG have orthogonal linear polarizations. 
The inset shows a part of the first BZ. 
(e)
Numerically obtained photonic band structure of the unit cell with $d_1$ = $d_2$ = 0.375$a$ for the lowest six bands. 
}
\label{fig:SCLband}
\end{figure}

%%%%%%%%%%%%%%%%%%%%%%%%%%%%%%%%%%%%%%%%%%%%%%%%%%%%%%%%%%%%%%%%%%%%%%%%%%%%%%%

We begin with a simple cubic lattice with a unit cell composed of dielectrics and air, as shown in Fig. 1(a). 
The refractive index of the dielectrics was 3.5. 
The period of the unit cell was $a$ = 12.8 mm, and the width of the dielectrics in the unit cell was the same in all three directions, that is, $w$/2 = 0.125$a$ = 1.6 mm. 
We then modified the structure by shifting the dielectric bars at the edges towards the center in the $x$ and $y$-directions, as shown in Fig. 1(b). 
The length of the shift was $d_1$/2, and $d_2$ satisfied $a$ = $d_1+d_2+w$. 
When $d_1$ = $a-w$ and $d_2$ = 0, the dielectric bars reached the center of the $x$-$y$ plane, as shown in Fig. 1(c). 
Here, the two types of unit cells in Figs. 1(a) and (c) are identical, except for the shifts of $a$/2 in the $x$- and $y$-directions. 
We labelled the unit cells in Figs. 1(a) and (c) as 'Type-I' and 'Type-II,' respectively. 

%%%%%%%%%%%%%%%%%%%%%%%%%%%%%%%%%%%%%%%%%%%%%%%%%%%%%%%%%%%%%%%%%%%%%%%%%%%%%%%

For these unit cells, we calculated the photonic band structures using the plane-wave expansion method. 
Because the difference between type-I and type-II was simply the shift in the origin of the unit cell, type-I and type-II had the same photonic band structure, as shown in Fig. 1(d). 
A cPBG was formed \cite{SCL} at a normalized frequency of approximately 0.41, corresponding to 9.5 GHz for a structure with $a$ = 12.8 mm. 
Below the cPBG, the two modes degenerated in the $\it{\Gamma}$-$Y$ and $\it{\Gamma}$-$R$ regions. 
The magnetic field distributions show that the lowest (red) and second-lowest (blue) modes at the $M_{YZ}$ point had linear polarizations parallel and perpendicular to the $y$-$z$ plane, respectively. 
The polarization direction of the lowest mode was tilted $-45^\circ$ from the $y$-axis. 
When $d_1$ = $d_2$ = ($a-w$)/2 = 0.375$a$, the bandgap disappeared, and the lowest four bands were almost compltetely degenerated in the $Y$-$M_{YZ}$ and $M_{YZ}$-$R$ regions, as shown in Fig. 1(e). 

%%%%%%%%%%%%%%%%%%%%%%%%%%%%%%%%%%%%%%%%%%%%%%%%%%%%%%%%%%%%%%%%%%%%%%%%%%%%%%%

Here, we performed theoretical and numerical calculations of the Zak phases in 3D PhCs. 
The Zak phase, also known as the 1D Berry phase or the Z$_2$ Berry phase, is a geometric phase that the eigenstate of the $n$-th band acquires during adiabatic evolution along a closed path in momentum space. 
In electronic systems, it represents the charge polarization in real space. 
This is mathematically defined as the 1D integral of the Berry connection over the first Brillouin zone (BZ) ~\cite{Zak}. 
In 1D multi-band systems, the Zak phase is determined for each sub-band ~\cite{Gao}. 
In 2D systems, the Zak phase can be determined in any in-plane directions ~\cite{Ota,Paz}. 
To generalize this notation to 3D systems, we numerically calculated the Zak phases in the 3D PhCs from the spatial distributions of electromagnetic fields obtained by using the finite-difference time-domain (FDTD) method. 
In general, the Berry phase of the $n$-th band was obtained by using the following closed path integral: 
\begin{equation} 
\label{zak}
Z^n = \oint_{BZ} {\bf A}^n\left({\bm{k}}\right) d {\bm{k}},
\end{equation}
where ${\bf A}^n\left({\bm{k}}\right)$ is the Berry connection, defined as
\begin{equation} 
\label{berryconnection}
{\bf A}^n\left({\bm{k}}\right) = i \braket{u^n_{\bm{k}}({\bm{r}})| \nabla_{\bm{k}} | u^n_{\bm{k}}({\bm{r}})},
\end{equation}
where $u^n_{\bm{k}}({\bm{r}})$ is the periodic part of the Bloch wave function. 

%%%%%%%%%%%%%%%%%%%%%%%%%%%%%%%%%%%%%%%%%%%%%%%%%%%%%%%%%%%%%%%%%%%%%%%%%%%%%%%

Because the Zak phase is a 1D integral in the first BZ, in any 3D system, it depends on a 2D plane perpendicular to the integral direction. 
Here, we show the calculation for the Zak phases in the $k_y$-$k_z$ plane, where the integral is taken along the $k_x$ direction. 
This method can be applied to calculate the Zak phases on other planes, such as the $k_x$-$k_y$ and $k_x$-$k_z$ planes. 
Equation~(\ref{zak}) can be rewritten as follows: 
\begin{equation} 
\label{zak1}
Z^n \left(k_y,k_z\right) = \int_{-\pi}^{\pi}
\braket{u^n_{k_x,k_y,k_z}\left({\bm{r}}\right) |i
\partial_{k_x}|u^n_{k_x,k_y,k_z}\left({\bm{r}}\right) } dk_x. 
\end{equation}
The integral of the Berry connection over $-\pi \leq k_x \leq \pi$ can be obtained by dividing the first BZ into small segments and approximating the integral by summing the contributions of each
segment. 
Further detailed calculations, including the overlap matrices for the degenerated modes, are presented in \cite{EPAPS-Takahashi}. 

%%%%%%%%%%%%%%%%%%%%%%%%%%%%%%%%%%%%%%%%%%%%%%%%%%%%%%%%%%%%%%%%%%%%%%%%%%%%%%%

The Zak phases for the type-I and type-II structures are $0$ or $\pi$, owing to the mirror or inversion symmetry in the unit cells. 
We calculated the Zak phases for the two bands below the cPBG in Fig. 1(d). 
The top and bottom rows in Fig. 2 show the numerical results of the Zak phases in the $k_y$-$k_z$ plane for the type-I and type-II structures, respectively. 
The $k_y$- and $k_z$-axes are considered in the first BZ: $-\pi/a \leq k_y \leq \pi/a$ and $-\pi/a \leq k_z \leq \pi/a$. 
For the type-I structure, the Zak phases were $\pi$ for the first lowest band and $0$ for the second lowest band at all points in the $k_y$-$k_z$ plane. 
For the type-II structure, the values were $0$ and $\pi$ for the first and second lowest bands, respectively, at all points in the $k_y$-$k_z$ plane. 
Therefore, the two types of unit cells were confirmed to exhibit opposing topological properties. 
Considering the continuous structural modification by varying $d_1$ in Fig. 1(a)--(c), we observed band inversions and topological transitions at $d_1$ = $d_2$ = 0.375$a$ in Fig. 1(e). 

%%%%%%%%%%%%%%%%%%%%%%%%%%%%%%%%%%%%%%%%%%%%%%%%%%%%%%%%%%%%%%%%%%%%%%%%%%%%%%%

\begin{figure}
\centering
\includegraphics[width=8cm]{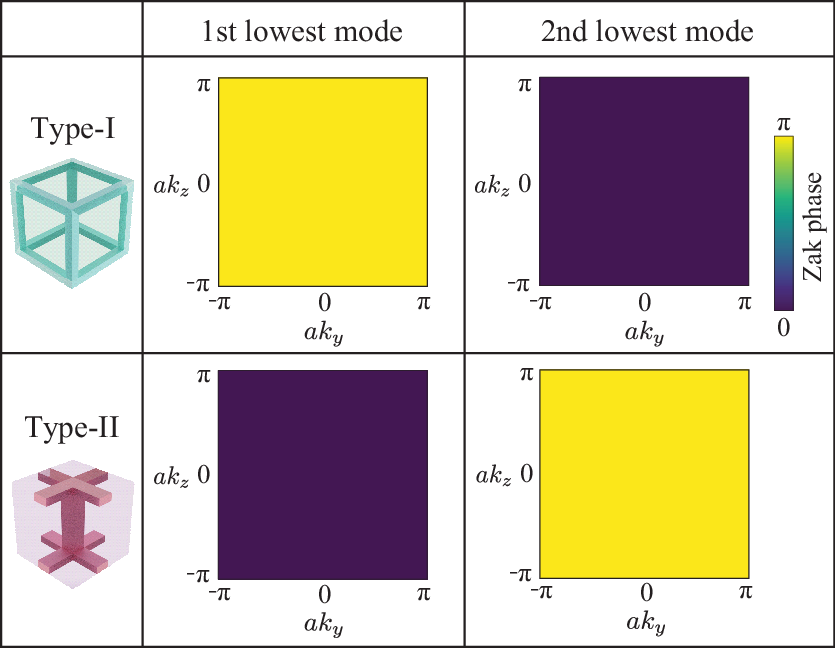}
\caption{
Numerically calculated Zak phases for the first and second lowest modes in Fig. 1(d). 
The top and bottom row shows the Zak phases in the $k_y$-$k_z$ plane for the type-I and type-II structures, respectively. 
The yellow and purple colors indicate that the Zak phases are all $\pi$ and $0$, respectively.
}
\label{zakphase}
\end{figure}

%%%%%%%%%%%%%%%%%%%%%%%%%%%%%%%%%%%%%%%%%%%%%%%%%%%%%%%%%%%%%%%%%%%%%%%%%%%%%%%

\begin{figure}
\centering
\includegraphics[width=7cm]{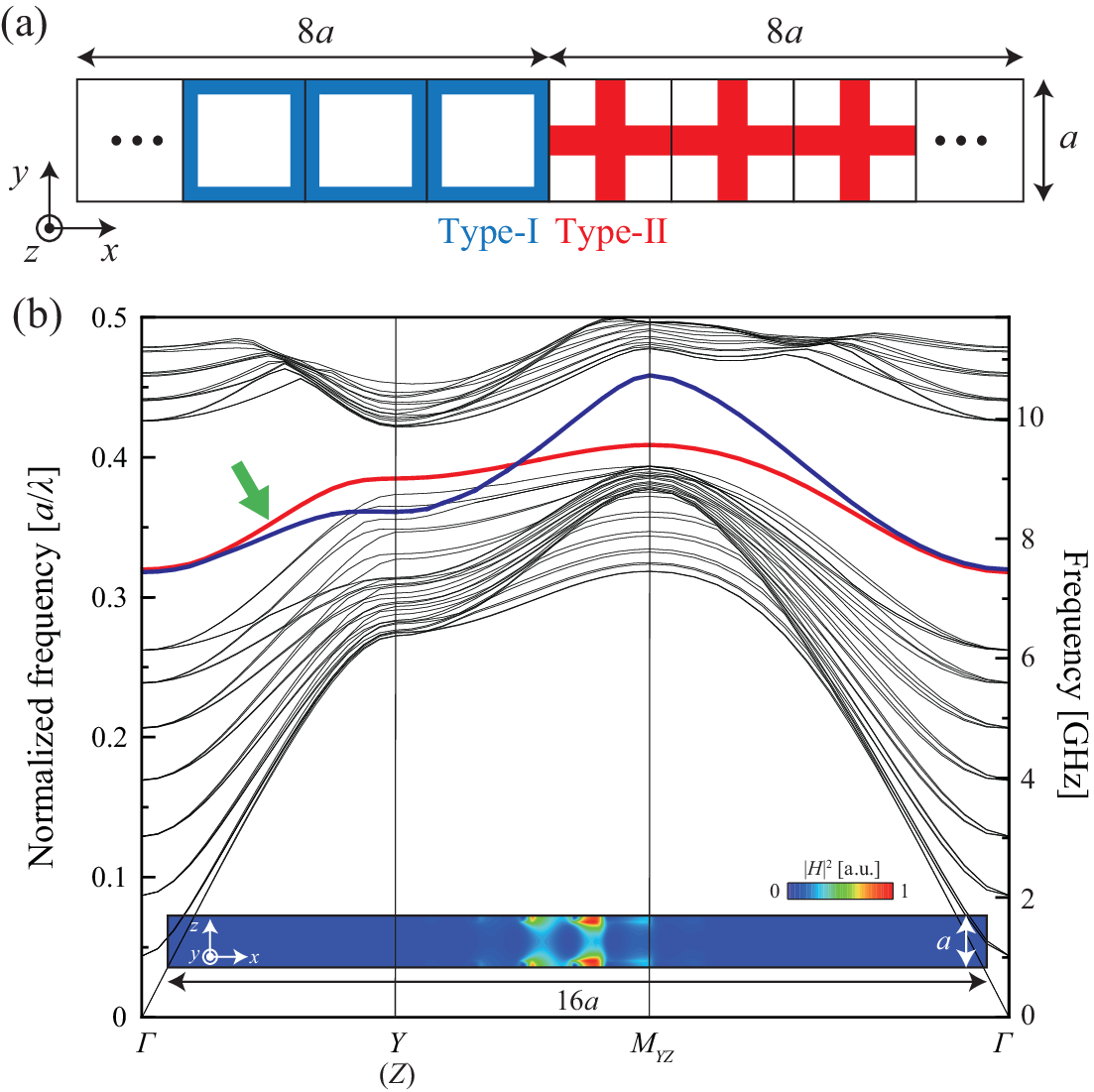}
\caption{
(a)
Schematic top view of the supercell for the calculation of 2D boundary states. 
The length of the supercell in the $z$-direction was $a$. 
(b)
Projected photonic band structure for the supercell in (a). 
The two boundary modes are in red and blue. 
The inset shows the numerically obtained spatial distribution of $|H|^2$ for the red-colored mode at the wavevector indicated by the green arrow in the main panel. 
The spatial map on the surface of the supercell in the $y$-direction was plotted. 
}
\label{fig:boundary_band}
\end{figure}

%%%%%%%%%%%%%%%%%%%%%%%%%%%%%%%%%%%%%%%%%%%%%%%%%%%%%%%%%%%%%%%%%%%%%%%%%%%%%%%

Owing to the Zak phase difference in the $x$-direction, the (first-order) topological boundary states were expected to form at the interface between the type-I and type-II structures. 
To investigate the dispersion curves of the boundary states, we calculated the projected band structure for a supercell containing type-I and type-II unit cells, as shown in Fig. 3(a), using the plane-wave expansion method. 
The lengths of the supercell were 16$a$, $a$, and $a$ in the $x$-, $y$-, and $z$-directions, respectively. 
In the $x$-direction, the eight left cells (type-I) and the eight right cells (type-II) formed an interface at the center. 

%%%%%%%%%%%%%%%%%%%%%%%%%%%%%%%%%%%%%%%%%%%%%%%%%%%%%%%%%%%%%%%%%%%%%%%%%%%%%%%

Figure 3(b) shows the projected photonic band structure of the supercell. 
Two additional modes, colored red and blue, appeared around the cPBG. 
The inset of Fig. 3(b) shows the spatial distribution of $|H|^2$ for the red-colored mode at the wavevector indicated by the green arrow in the main panel. 
This field distribution was obtained using an FDTD method with perfectly matched layers as the boundary conditions in the $x$-direction to extract the boundary state. 
The details of the calculation setup are described in \cite{EPAPS-Takahashi}. 
The spatial distribution on the surface of the supercell in the $y$-direction was plotted. 
The localized magnetic-field intensity at the interface confirmed the boundary state where the electromagnetic field propagates in the 2D plane of the interface.

%%%%%%%%%%%%%%%%%%%%%%%%%%%%%%%%%%%%%%%%%%%%%%%%%%%%%%%%%%%%%%%%%%%%%%%%%%%%%%%

The red-colored boundary mode at the $M_{YZ}$ point mainly had $H_y$ and $H_z$ components with comparable amplitudes, indicating a linear polarization parallel to the $y$-$z$ plane and showing that the boundary mode was derived from the lowest bulk mode with the same polarization at the $M_{YZ}$ point. 
For the blue mode in Fig. 3(b), $|H|^2$ was localized at the interface \cite{EPAPS-Takahashi}. 
The dominant magnetic component of the mode was $H_x$, indicating that the blue-colored boundary mode was derived from the second-lowest bulk mode polarized in the $x$-direction. 
Therefore, the two boundary modes were caused by distinct Zak phases in each bulk mode below the bandgap.

%%%%%%%%%%%%%%%%%%%%%%%%%%%%%%%%%%%%%%%%%%%%%%%%%%%%%%%%%%%%%%%%%%%%%%%%%%%%%%%

These boundary modes were also obtained at the interface between unit cells with $d_1$ = 0.15$a$ and 0.6$a$, where the Zak phases were different. 
In contrast, there was no boundary mode at the interface between the unit cells with $d_1$ = 0 and 0.15$a$ or $d_1$ = 0.65$a$ and 0.75$a$, where the Zak phases were the same \cite{EPAPS-Takahashi}. 
These results confirm that the difference in the Zak phase caused topological boundary states and that topological transition was not induced by the small perturbation of $d_1$. 

%%%%%%%%%%%%%%%%%%%%%%%%%%%%%%%%%%%%%%%%%%%%%%%%%%%%%%%%%%%%%%%%%%%%%%%%%%%%%%%

Because the $x$- and $y$-directions are equivalent in both type-I and type-II, the $\pi$ difference in the Zak phases between type-I and type-II also occurred in the $y$-direction. 
Therefore, second-order topological states were expected to form at the corners of the interfaces \cite{Xie,Liu,Liu2} when a type-I structure is surrounded by a type-II structure, as schematically shown in Fig. 4(a). 
To investigate the dispersion curves of the hinge states, we calculated the projected band structure for the supercell (Fig. 4(a)) using the finite-element method. 
The lengths of the supercell were 12$a$, 12$a$, and $a$ in the $x$-, $y$-, and $z$-directions, respectively, and the type-II structure was a 6$a$ square placed at the center. 
Figure 4(b) shows the projected band structure of the supercell. 
Among the many modes folded in the $\it{\Gamma}$-$Z$ region, we separated the bulk modes colored in gray, boundary modes in blue, and hinge modes in red by calculating the spatial distributions of the magnetic field intensity for each mode. 
We found that the four almost degenerate modes, corresponding to the number of corners in Fig. 4(a), appeared around the cPBG, and that $|H|^2$ were spatially localized at the corners. 

%%%%%%%%%%%%%%%%%%%%%%%%%%%%%%%%%%%%%%%%%%%%%%%%%%%%%%%%%%%%%%%%%%%%%%%%%%%%%%%

\begin{figure}
\centering
\includegraphics[width=8cm]{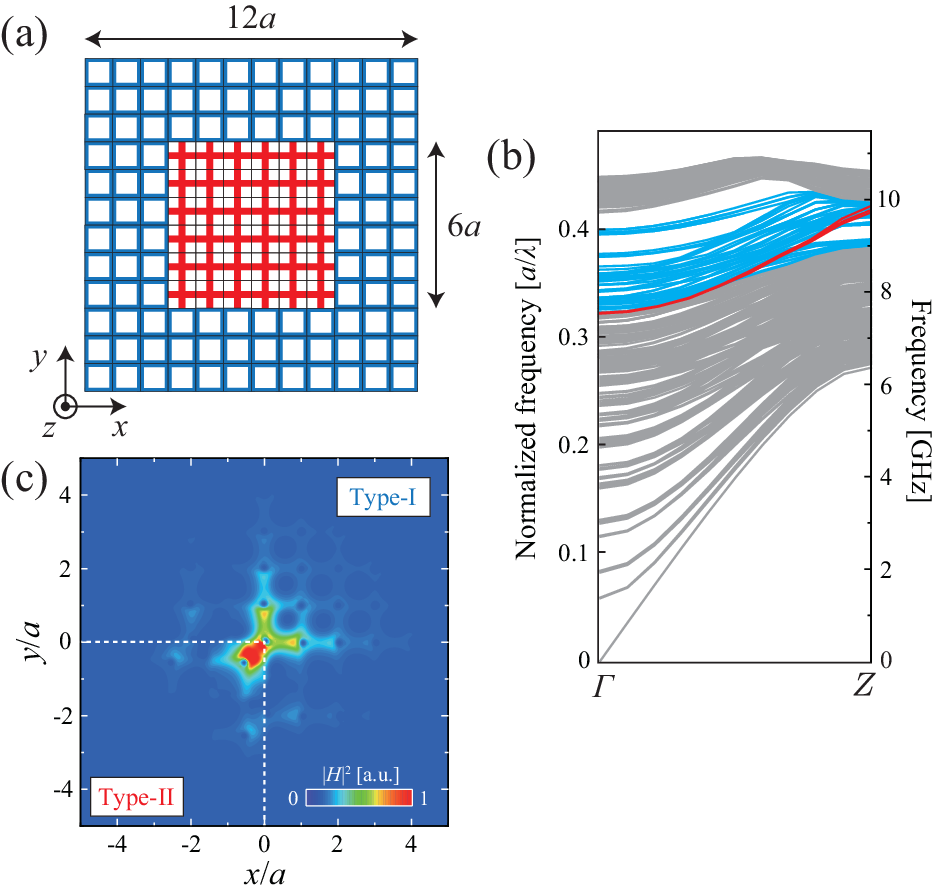}
\caption{
(a)
Schematic top view of the supercell for the calculation of 1D hinge states. 
Length of the supercell in the $z$-direction is $a$. 
The type-II structure is surrounded by the type-I structure, forming four corners. 
(b)
Projected photonic band structure for the supercell in (a). 
The bulk, boundary, and hinge modes are gray, blue, and red, respectively. 
(c)
Spatial distribution of $|H|^2$ at a normalized frequency of 0.315. 
The spatial map on the surface of the supercell in the $z$-direction was plotted. 
The magnetic field intensity was localized at the corner indicated by the white dotted lines. 
}
\label{fig:hinge_band_PRB}
\end{figure}

%%%%%%%%%%%%%%%%%%%%%%%%%%%%%%%%%%%%%%%%%%%%%%%%%%%%%%%%%%%%%%%%%%%%%%%%%%%%%%%

We also calculated the spatial distribution of $|H|^2$ at one of the four corners using the FDTD method. 
We adopted perfectly matched layers as the boundary conditions in the $x$- and $y$-directions to extract a single hinge state \cite{EPAPS-Takahashi}. 
Figure 4(c) shows the obtained spatial map of $|H|^2$ on the surface of the unit cell in the $z$-direction for a normalized frequency of 0.315 at the $\it{\Gamma}$ point. 
The magnetic field intensity was clearly localized at the corner indicated by the white dotted lines, and the maximum intensity was slightly shifted in the ($-x$, $-y$) direction. 
This hinge state mainly showed linear polarization, which was parallel to the $x$-$y$ plane and tilted $-45^\circ$ from the $x$-axis. 
These numerical results confirm that the hinge state was formed in a 3D structure based on simple cubic lattices. 
The obtained hinge state was derived from one of the boundary modes that stemmed hierarchically from the corresponding bulk mode in Fig. 1(d). 
This topological hierarchy across dimensions was based on a higher-order of bulk-boundary correspondence \cite{Benalcazar}. 
Similarly, another hinge mode derived from another boundary mode that hierarchically stems from the corresponding bulk mode in Fig. 1(d) may also exist. 
However, we did not find another hinge mode, probably because this mode lies away from the cPBG and was largely overlapped by the bulk and boundary modes. 

%%%%%%%%%%%%%%%%%%%%%%%%%%%%%%%%%%%%%%%%%%%%%%%%%%%%%%%%%%%%%%%%%%%%%%%%%%%%%%%

\section{Sample fabrication and measurement setups}

%%%%%%%%%%%%%%%%%%%%%%%%%%%%%%%%%%%%%%%%%%%%%%%%%%%%%%%%%%%%%%%%%%%%%%%%%%%%%%%

\begin{figure}
\centering
\includegraphics[width=8.5cm]{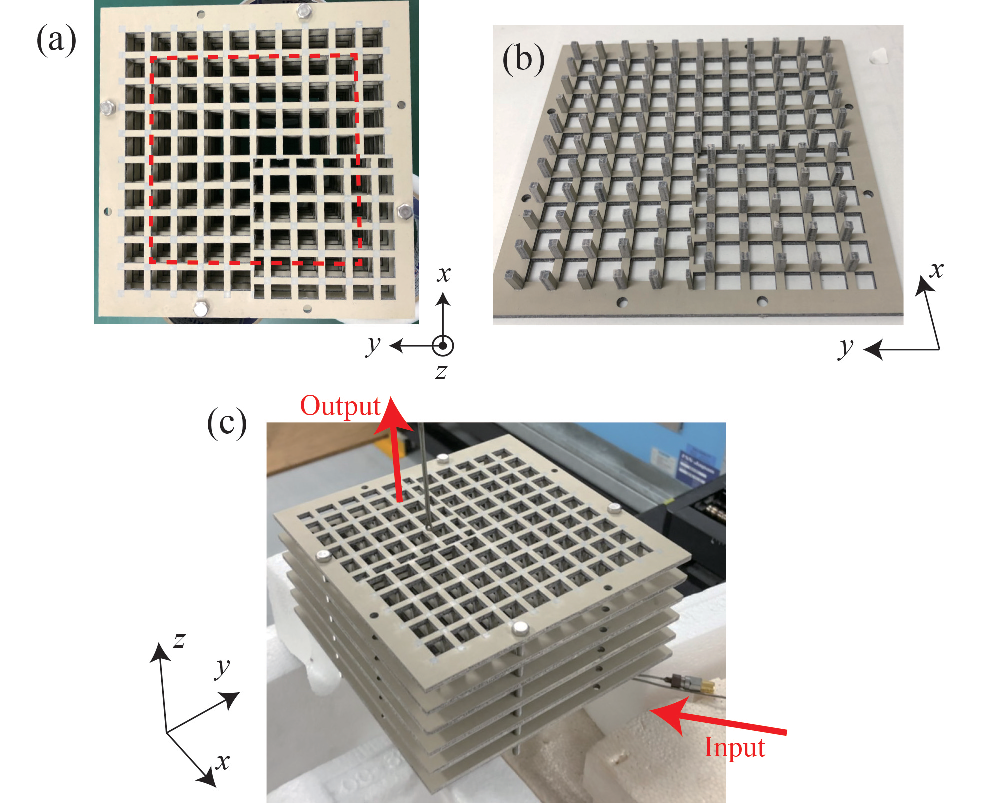}
\caption{
(a)
Photo of the fabricated 3D PhC from the top view. 
The 100 mm square area indicated by the red dotted square was scanned by the detection antenna in (c). 
(b)
Photo of a fabricated single period in the $z$-direction. 
Two kinds of pillars were glued at the designed positions on the patterned board. 
(c)
Photo of the measurement setup around the PhC. 
The incident transmission line was inserted from the direction tilted by $+45^\circ$ from the $x$-axis, and the loop antenna terminating the transmission line worked as a magnetic dipole oscillating in the direction perpendicular to the transmission line and $z$-axis. 
The incident loop antenna was touched on the bottom surface at the center of the pattern. 
The detection loop antenna was touched on the top surface and mechanically scanned in the $x$- and $y$-directions. 
The directions of the two loop antennas were parallel to each other. 
During the measurement, absorber sheets were attached on the side surfaces of the structure. 
}
\label{fig:sample_reduced}
\end{figure}

%%%%%%%%%%%%%%%%%%%%%%%%%%%%%%%%%%%%%%%%%%%%%%%%%%%%%%%%%%%%%%%%%%%%%%%%%%%%%%%

Based on the numerical results, we fabricated a 3D structure using polytetrafluoroethylene boards with a thickness of 3.2 mm, refractive index of 3.5 ($\pm$0.1), and dissipation factor of $\sim$0.002 at 10 GHz. 
First, we drilled dielectric boards and formed a 2D pattern from the top view, as shown in Fig. 5(a). 
The dimensions of the 2D pattern were identical to those of the designed structure. 
The number of in-plane periods was 10, and the pattern was surrounded by a frame that had small holes with a diameter of 3.3 mm to align the following stack in the $z$-direction. 
Hundreds of square pillars and tens of rectangular pillars were prepared by cutting the boards. 
The bottom and height of the square pillars were a 3.2 mm square and 9.6 mm, respectively. 
The bottom and height of the rectangle pillars were a 1.6 mm $\times$ 3.2 mm rectangle and 9.6 mm, respectively. 
Finally, we individually glued the pillars at the designed positions on the patterned boards, as shown in Fig. 5(b), and constructed a 3D structure by stacking five periods in the $z$-direction. 
The final 3D structure was a 152 mm in-plane square with a height of 67.2 mm. 
The fabrication error was $\pm$0.2 mm in the in-plane direction and $\pm$0.5 mm in the $z$-direction, which are much smaller than the target wavelength of $\sim$30 mm (10 GHz). 
The square pillars were glued at the center of the pattern instead of the chipped square pillars because of fabrication limitations, which were numerically confirmed to provide little change in the hinge states.

%%%%%%%%%%%%%%%%%%%%%%%%%%%%%%%%%%%%%%%%%%%%%%%%%%%%%%%%%%%%%%%%%%%%%%%%%%%%%%%

For the fabricated PhC shown in Fig. 5(c), we performed microwave transmission measurements in the $z$-direction using a network analyzer (N5224A, Agilent Technologies). 
As an incident magnetic dipole, a loop antenna with a diameter of 3 mm was placed on the bottom surface at the center of the $x$-$y$ plane. 
The direction of the magnetic dipole was parallel to the $x$-$y$ plane and tilted $-45^\circ$ from the $x$-axis along the linear polarization of the hinge state. 
To measure the spatial distribution of the transmitted microwaves on the top surface of the 3D structure, another loop antenna was placed on the top surface and mechanically scanned in the $x$-$y$ plane with a spatial resolution of 2 mm. 
The scanned area was a 100 mm square that shared the same center as the pattern, as indicated by the red dotted square in Fig. 5(a). 
Thus, the total number of measured points was 51 $\times$ 51. 
The detection loop antenna was also tilted by $-45^\circ$ from the $x$-axis, parallel to the incident loop antenna. 
From the network analyzer, we obtained one of the scattering parameters, $S_{21}$, between the incident loop antenna (port 1) and the detection loop antenna (port 2) in the frequency range of 6.50--11.30 GHz with a resolution of 0.06 GHz. 
During the measurement, we attached absorber sheets with an absorption of 23 dB at 10 GHz on the side surfaces of the PhC. 

%%%%%%%%%%%%%%%%%%%%%%%%%%%%%%%%%%%%%%%%%%%%%%%%%%%%%%%%%%%%%%%%%%%%%%%%%%%%%%%

\section{Experimental results and discussion}

%%%%%%%%%%%%%%%%%%%%%%%%%%%%%%%%%%%%%%%%%%%%%%%%%%%%%%%%%%%%%%%%%%%%%%%%%%%%%%%

\begin{figure}
\centering
\includegraphics[width=8cm]{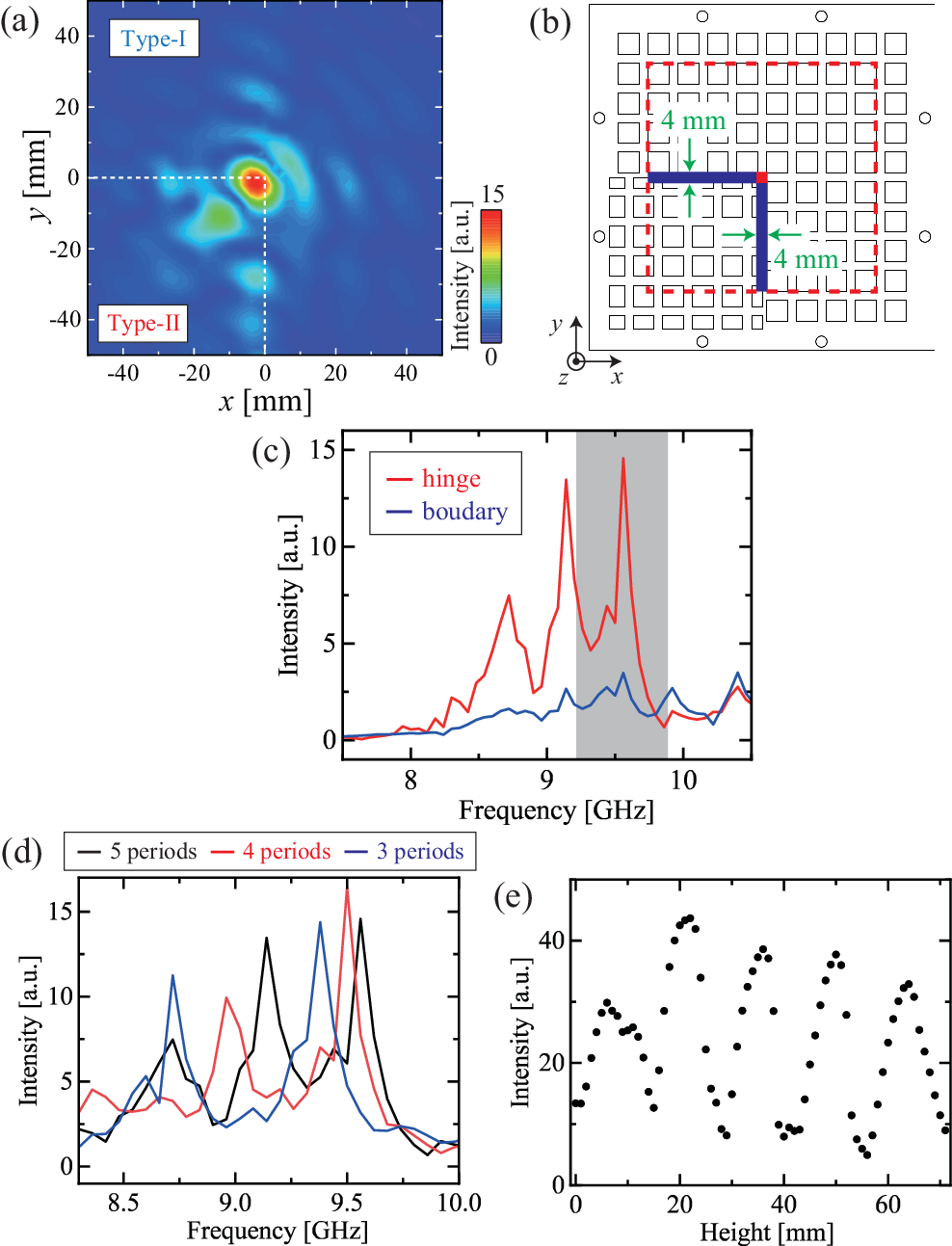}
\caption{
(a)
Measured spatial distribution of transmission intensity at 9.56 GHz. 
Localized intensity at the corner shown by the white dotted lines indicates the hinge state. 
(b)
Schematic of the defined area as the corner (red) and the boundary (blue). 
The measured transmission intensity in these areas was extracted to plot the transmission spectra for each area in (c). 
The red dotted square indicates the scanned 100 mm square, which is equivalent to that in Fig. 5(a). 
(c)
Transmission spectra of the corner (red) and the boundary (blue) areas represent the hinge and the boundary states, respectively. 
The intensity was averaged in each area in (b). 
The shaded region indicates the cPBG where the hinge state shows the largest intensity. 
(d)
Measured transmission spectra for the PhC having different numbers of periods in the $z$-direction. 
The black curve of 5 periods is equivalent to the red curve in (c). 
(e)
Transmission intensity as a function of the height of the detection antenna, which was inserted into the smallest square hole around the center of the in-plane pattern. 
The height of 0 mm indicates the bottom of the PhC. 
}
\label{fig:hinge_exp}
\end{figure}

%%%%%%%%%%%%%%%%%%%%%%%%%%%%%%%%%%%%%%%%%%%%%%%%%%%%%%%%%%%%%%%%%%%%%%%%%%%%%%%

Figure 6(a) shows the measured spatial distribution of the transmission intensity on the top surface at 9.56 GHz. 
The intensity was localized at the center, indicating a hinge state. 
Because the bulk and boundary modes were absent in the $\it{\Gamma}$-$Z$ region at approximately 9.56 GHz in Fig. 3(b), a strong intensity localized at the center was observed despite the co-existing boundary modes in Fig. 4(b). 
This is also because we selected the linear polarization parallel to the hinge state by fixing the angle of the loop antennas. 
The maximum intensity shifted slightly in the ($-x$, $-y$) direction, which is consistent with the numerical results in Fig. 4(c). 
We also found clear intensity localization around the boundaries for a boundary state \cite{EPAPS-Takahashi}. 

%%%%%%%%%%%%%%%%%%%%%%%%%%%%%%%%%%%%%%%%%%%%%%%%%%%%%%%%%%%%%%%%%%%%%%%%%%%%%%%

To investigate the frequency dependence of the transmission intensity at the boundary and corner, we defined the areas of the boundary and corner separately and extracted the measured intensity in each area at each frequency. 
Figure 6(b) shows the corner (red) and boundary (blue) areas. 
The red-colored area at the center of the pattern is a 4-mm square containing 4 data points. 
By averaging the intensity data, we assigned the transmission intensity to the hinge state. 
The blue-colored areas along the boundaries are composed of two 4 mm $\times$ 48 mm rectangles containing 96 data points. 
By averaging the intensity data, we assigned the transmission intensity to the boundary states. 
Figure 6(c) shows the transmission spectra obtained for the boundary and hinge states. 
The high-intensity frequency region for each state overlapped with that of the numerically obtained modes in Figs. 3(b) and 4(b). 
In the region from 8.5 to 9.8 GHz, the intensity of the hinge state was much larger than that of the boundary states, suggesting that the incident microwave mainly coupled to the hinge state. 
In addition, the hinge state showed a maximum intensity at 9.56 GHz, which was inside the cPBG, as indicated by the shaded region in Fig. 6(c). 
This is because there was ideally no leakage to the bulk modes. 
In the low-frequency region below 8 GHz, the intensities of both modes were small, although both modes had dispersions in the region, as shown in Figs. 3(b) and 4(b). 
This absence of intensity was probably caused by a large overlap with the bulk modes. 

%%%%%%%%%%%%%%%%%%%%%%%%%%%%%%%%%%%%%%%%%%%%%%%%%%%%%%%%%%%%%%%%%%%%%%%%%%%%%%%

For the hinge state, several intense peaks appeared at 8.72, 9.14, and 9.56 GHz. 
These peaks were caused by Fabry--P\'{e}rot resonances in the hinge waveguide owing to the air interfaces at the top and bottom of the structure. 
To verify the Fabry--P\'{e}rot resonances experimentally, we measured the transmission intensity of the structures with different numbers of periods in the $z$-direction. 
When the stacking period was reduced, the peak frequencies shifted, as shown in Fig. 6(d). 
The experimentally obtained free spectral ranges (FSRs) were 1.44, 1.90, and 2.42 mm for five, four, and three stacking periods, respectively. 
These results were almost the same as 1.77 and 2.40 mm for the four and three stacking periods, respectively, which were calculated by using the 1.44 mm FSR for the five periods and reducing the structural height. 
In addition, from the FSR for the five stacking periods, we obtained a group index of 5.5, which is consistent with the numerical results obtained from the dispersion of the hinge mode \cite{EPAPS-Takahashi}. 

%%%%%%%%%%%%%%%%%%%%%%%%%%%%%%%%%%%%%%%%%%%%%%%%%%%%%%%%%%%%%%%%%%%%%%%%%%%%%%%

Furthermore, we measured the spatial distribution of the transmission intensity inside the PhC for the hinge state. 
We inserted the detection antenna into the smallest square hole around the center of the in-plane pattern and scanned it in the $z$-direction. 
Figure 6(e) shows the measured intensity as a function of the height of the detection antenna, showing the interference fringes. 
In Fig. 6(e), a height of 0 mm indicates the bottom of the PhC, and the height of the five-period PhC is 67.2 mm. 
Using a Fourier transform of the fringes, we obtained a spatial wavenumber of 0.215, which is almost consistent with the wavenumber of 0.95 $\times$ $\pi/a$ = 0.233 at 9.56 GHz in the band structure shown in Fig. 4(b). 
The slight deviation in the wavenumber was caused by the small number of samples, as shown in Fig. 6(e). 
From these experimental and numerical results, we confirmed that the microwave propagation through the hinge state was successfully observed in the studied 3D structure based on simple cubic lattices. 

%%%%%%%%%%%%%%%%%%%%%%%%%%%%%%%%%%%%%%%%%%%%%%%%%%%%%%%%%%%%%%%%%%%%%%%%%%%%%%%

\section{Summary}

%%%%%%%%%%%%%%%%%%%%%%%%%%%%%%%%%%%%%%%%%%%%%%%%%%%%%%%%%%%%%%%%%%%%%%%%%%%%%%%

We theoretically and numerically investigated a hinge state in a fabricated 3D PhC insulator composed of two types of cubic unit cells and experimentally confirmed the existence of a hinge state in the microwave region. 
The proposed simple dielectric structures in each unit cell, type-I and II, were identical, except for a shift of $a$/2 in the $x$- and $y$-directions. 
Both type-I and type-II PhCs had the same cPBG. 
Owing to the mirror or inversion symmetry in a simple cubic lattice, the Zak phases were used to determine the topological properties of the unit cells. 
We numerically confirmed the difference in the Zak phases between the type-I and type-II unit cells in the orthogonal directions for each band below the bandgap. 
When the type-II structure was surrounded by the type-I structure, boundary and hinge states appeared at the boundary and corner, respectively, according to the higher-order of bulk-boundary correspondence. 
Microwave measurements of the fabricated 3D PhC containing two boundaries and one corner showed localized intensity, confirming the boundary and hinge states. 
Our findings can be applied to the 3D control of microwave propagation for communication, and is an important step towards the realization of robust 3D photonic circuits for infrared light. 

%%%%%%%%%%%%%%%%%%%%%%%%%%%%%%%%%%%%%%%%%%%%%%%%%%%%%%%%%%%%%%%%%%%%%%%%%%%%%%%

\section{Acknowledgements}\label{sec:acknowledgements}

%%%%%%%%%%%%%%%%%%%%%%%%%%%%%%%%%%%%%%%%%%%%%%%%%%%%%%%%%%%%%%%%%%%%%%%%%%%%%%%

This study was supported by JSPS KAKENHI under Grant Numbers 17H06138, 18K18857, 21H01019, and JST-CREST JPMJCR19T1. 

%%%%%%%%%%%%%%%%%%%%%%%%%%%%%%%%%%%%%%%%%%%%%%%%%%%%%%%%%%%%%%%%%%%%%%%%%%%%%%%

%%%%%%%%%%%%%%%%%%%%%%%%%%%%%%%%%%%%%%%%%%%%%%%%%%%%%%%%%%%%%%%%%%%%%%%%%%%%%%%

\end{document}